\documentclass[11pt,twoside
]{article}

\usepackage{baaa2008}
\usepackage{graphicx}
\usepackage{subfigure}
\usepackage{psfrag}
\usepackage{amssymb}
\usepackage[spanish,activeacute,english]{babel}
\usepackage[latin1]{inputenc}
\usepackage[T1]{fontenc} 
\usepackage{ae,aecompl} 
\usepackage{latexsym}
\usepackage{verbatim}
\usepackage{amsmath}
\usepackage{amsfonts}
\usepackage{amssymb}
\usepackage{wasysym}
\usepackage[colorlinks=true,dvips]{hyperref}

\begin{document}
\myselectenglish
\vskip 1.0cm
\markboth{ V. Bosch-Ramon }%
{}

\pagestyle{myheadings}
\vspace*{0.5cm}
\parindent 0pt{TRABAJO INVITADO} 
\vskip 0.3cm
\title{Powerful non-thermal emission in black-hole powered sources}


\author{V. Bosch-Ramon$^{1}$}

\affil{%
 (1) Max Planck Institut f\"ur Kernphysik \\
}

\begin{abstract} 

Powerful non-thermal emission has been detected coming from relativistic collimated 
outflows launched in the vicinity of black holes of a very wide range of masses, from few to $\sim
10^{10}$~M$_{\odot}$. These collimated 
outflows or jets have large amounts of energy and momentum extracted from the black hole itself and/or from matter trapped in its potential well.
The key ingredients for the formation of these powerful jets
are accretion of matter with angular momentum, the huge gravitational potential of the compact object, the strong ordered magnetic fields
near the black-hole horizon, the potentially large rotational energy in the case of a Kerr black hole, and an escape velocity close to $c$. At different scales along 
the outflows,
i.e. $\sim 10-10^{10}\,R_{\rm Sch}$ from the black hole, the local conditions can lead to the generation of non-thermal populations of particles via, e.g., magnetic reconnection,
magneto-centrifugal mechanisms, diffusive processes, or the so-called
converter mechanism. These non-thermal populations of particles, interacting with dense matter, magnetic, and
radiation fields, could yield radio-to-gamma-ray emission via synchrotron process, inverse Compton scattering, relativistic Bremsstrahlung, proton-proton and photo-hadron
colissions, and even heavy nuclei photo-disintegration. Other processes, like pair creation or the development of electromagnetic cascades, could be also relevant in black-hole
jets and their surroundings. Black holes of different masses, accretion rates and environments show different phenomenologies, as can be observed in
AGNs, GRBs or microquasars. Nonetheless, these sources
basically share the same fundamental physics: accretion, black-hole rotation, plus an environment, but they are individualized due to their own specific conditions. 
In this paper, we
qualitatively review the main characteristics of the non-thermal emission produced in jets from black holes, giving also a brief overview on the physical properties of black
hole/jet systems. We comment as well on some important differences and similarities between classes of sources, and on the prospects for the study of the non-thermal
emission from astrophysical sources powered by black holes.

\end{abstract}

\begin{resumen}

En las cercan\'ias de agujeros negros de un amplio rango de masa, $\sim 1-10^{10}$~M$_{\odot}$, se eyectan poderosos chorros relativistas de plasma que producen emisi\'on no
t\'ermica. Estos chorros pueden transportar grandes cantidades de energ\'ia y momento extra\'idos del agujero negro y/o de la mater\'ia atrapada en su pozo gravitacional. Para la
formaci\'on de estos chorros de plasma, los ingredientes clave son la acreci\'on de materia con momento angular, el intenso potencial gravitacional del agujero negro, la presencia
de fuertes campos magn\'eticos ordenados, la posiblemente gran energ\'ia rotacional del agujero negro, y la alta velocidad de escape, que puede ser cercana a $c$. A diferentes
escalas espaciales de los chorros de plasma, a $\sim 10-10^{10}\,R_{\rm Sch}$ del agujero negro, las condiciones locales pueden llevar a la generaci\'on de poblaciones no t\'ermicas de
part\'iculas via reconexi\'on magn\'etica, un mecanismo magneto-centr\'ifugo, procesos difusivos, o el llamado mecanismo de conversi\'on. Estas poblaciones no t\'ermicas de
part\'iculas, interactuando con densos campos de radiaci\'on, magn\'eticos, y de materia, dan lugar a emisi\'on de fotones de muy diferentes energ\'ias por radiaci\'on
sincrotr\'on, dispersi\'on Compton inverso, Bremsstrahlung relativista, colisiones prot\'on-prot\'on, y foto-desintegraci\'on de n\'ucleos at\'omicos pesados. Otros procesos, como
creaci\'on de pares electr\'on-positr\'on, o el desarrollo de cascadas electromagn\'eticas, podr\'ian ser tambi\'en relevantes en estos chorros y en sus inmediaciones. Agujeros
negros de diferentes masas y medios circundantes mostrar\'an diferentes fenomenolog\'ias, como occure con los n\'ucleos de galaxias activas, las explosiones de rayos gamma, o los
microcu\'asares, que son fuentes que comparten b\'asicamente la misma f\'isica fundamental: acreci\'on, agujero negro en rotaci\'on, y un medio circundante, aunque con importantes
diferencias espec\'ificas. En este trabajo, se discuten a nivel cualitativo las caracter\'isticas m\'as significativas de la emisi\'on no t\'ermica proveniente de chorros lanzados
por agujeros negros, y se resumen brevemente las propiedades f\'isicas de este tipo de objetos. Tambi\'en se consideran las principales diferencias y similitudes entre clases de
fuentes, y se comentan las perspectivas del estudio de la emisi\'on no t\'ermica en fuentes con agujeros negros. 

\end{resumen}

\section{Introduction}

There is very strong observational evidence for the existence of black holes. From the epoch of the discovery of the X-ray source  Cygnus~X-1 as a binary system likely harboring a
black hole (Bolton 1972; Webster \& Murdin 1972), to the present days,  the number of black hole candidates has grown substantially at stellar as well as at super massive scales
(e.g. Casares 2001; Sch\"odel et al. 2002). In general, among other methods, evidence for a black hole comes from observations giving dynamical information
on the mass of the object under study, since compact objects more massive than black holes are not expected for certain masses ($M>3$~M$_\odot$ for stellar mass objects; e.g.
Casares 2001; see also Paredes 2008a). Lower mass black holes could also exist, although to find them we are constraint by our present knowledge on the physics of the collapsed object material.
Hereafter, we will assume that black holes do exist, and discuss how they could be involved in very energetic and luminous events in the Universe. We will focus our study on the
production of non-thermal radiation, which is strongly linked to the generation of jets.

The deep potential well of black holes can allow to release large amounts  of potential and rotational energy to the infinity (e.g. Lynden-Bell 1969; see also Salpeter 1964).
Provided that a particle falling onto the black hole can reach very small distances from the center of gravity  before reaching the event horizon, an amount equivalent to a
significant fraction of the particle rest-mass energy can be converted into kinetic energy. In the case of a gas, the situation is more complicated given collective effects,
although the energy of the infalling material can be partially tapped in the form of relativistic outflows or jets of plasma, which would carry a substantial fraction of the
energy and angular momentum of the incoming matter (e.g. Blandford 1976). Another energy provider of jets would be the rotational energy of the black hole, which may be extracted
via a Penrose-like (Penrose \& Floyd 1971) mechanism: some material in the black-hole ergosphere acquires negative energy for an observer in the infinity; this implies that some
material gets positive energy (as seen from infinity) and escapes far from the black-hole potential well.

Since several decades ago (e.g. Burbidge 1956), it has been known that non-thermal emission is generated in the jets produced in compact objects. In these jets, non-thermal
populations of particles (or cosmic rays -CRs-)  are generated and can interact with dense matter, magnetic, and radiation fields. Leptonic and hadronic processes associated with
these interactions yield radiation from radio to gamma-rays, and also neutrinos. The electromagnetic emission, as well as neutrinos and CRs, can be  observed with instruments,
either ground or space based, working in very different spectral domains, covering  almost 30 orders of magnitude in energy, from $\sim 10^{-6}$ to 10$^{22}$~eV (i.e. from radio
up to ultra high-energies -UHE-). There are basically three known classes of sources that harbor black holes and are powerful non-thermal emitters: Active Galactic Nuclei (AGNs),
Gamma-Ray Bursts (GRBs) and microquasars (MQs) (e.g. Begelman et al. 1984; M\'esz\'aros 2006; Mirabel \& Rodr\'iguez 1999). All these sources basically share the same fundamental
physics, although show distinct spectral, spatial and temporal behavior due to their own specific characteristics. 

In this paper, we qualitatively discuss the main features of the non-thermal emission of jets, ejected from the vicinity of black holes, trying to give at the same time a very
brief overview on the physical properties of these objects. We also comment on what are the main differences and similarities between classes of sources, and on the prospects for
the study of non-thermal emission in black-hole powered sources. It is not intended to discuss here neither observations nor theory of these objects, thus we refer to the cited
works where relevant observational and theoretical references can be found. Also, it is noted that the brief overview presented here does not intend to be exhaustive, but just
touch in a simple manner the different concerned topics.

\begin{figure}[!ht]
  \centering
  \includegraphics[width=.7\textwidth]{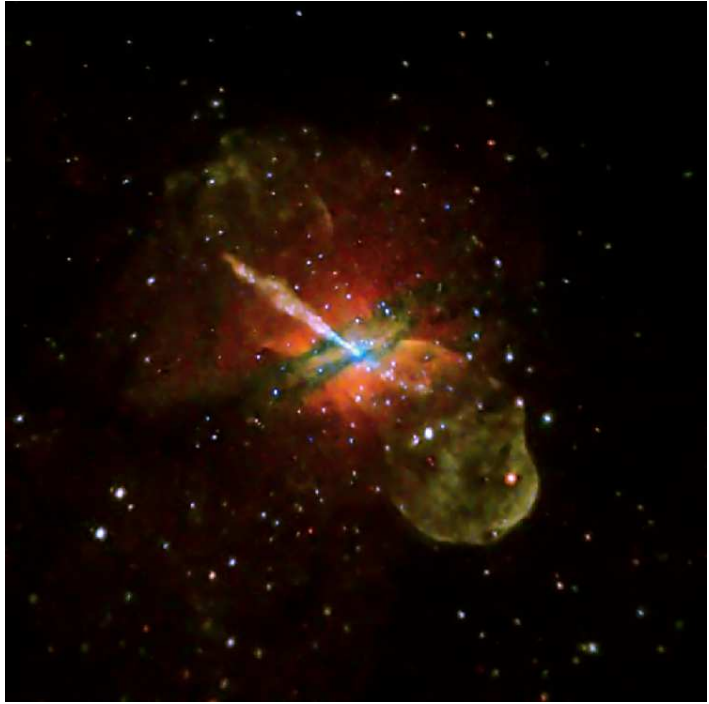}  
  \caption{View of the AGN Centaurus~A. The jet is clearly visible, as well as the structure formed by the interaction 
  of the jet with the external medium (credit: NASA/CXC/CfA/R.Kraft et al.). Centaurus~A may be one of the dominant sources 
  of UHE CRs (as proposed e.g. by Romero et al. 1996; see also recent AUGER results in Abraham et al. 2007).}
  \label{agn}
\end{figure}

For illustrative purposes, we present in Fig.~\ref{agn} an example of jets associated with a black-hole powered source.

\section{Non-thermal radiation from black holes}

\subsection{The black-hole engine}

Black holes are regions of the spacetime separated from the rest of the Universe
by an event horizon, which is a closed surface in which all the light cones
point towards inside the black hole. This implies that no matter, nor signal,
can propagate outwards from this surface. 
The existence of this surface
is caused by the presence of matter and fields inside the horizon with an
energy-momentum large enough as to produce such a distortion of the metric of
the space-time or, put in different words, to create such a large gravitational
potential that even massless particles cannot escape. Black holes are described
by the solutions of the Einstein equations of the general relativity (Einstein
1915) that represent a collapsed object (see Romero 2008a for a recent review).
In the context of general relativity, black holes are described by just three
quantities: mass, angular momentum, and charge, and the metric describing the
different types of black hole are the Schwartzchild's metric (massive;
Schwarzschild 1916a, 1916b); Kerr's metric (massive and rotating; Kerr 1963);
Reissner-Nordstr\"om's metric (massive and charged; Reissner 1916, and
Nordstr\"om 1918); and Kerr-Newmann's metric (massive, rotating and charged;
Newmann et al. 1965). As already mentioned and explained further in the
following sections, the deep gravitational potential of a black hole allows
accretion to be a very efficient energy source, and the same for the rapid
rotation of a Kerr black hole. The energy is finally released in the form of
electromagnetic radiation, CRs and neutrinos, making of black holes the powerful
engines of the brightest objects in the Universe. 

\subsection{Accretion and jet formation}

Accretion of matter can render the conditions for the generation of powerful jets. The energy required to launch these jets is extracted either from the accreted matter itself
(e.g. Blandford 1976; Lovelace 1976; Blandford \& Payne 1982), or from the rotation of the black hole (e.g. Blandford \& Znajek 1977; Punsly \& Coroniti 1990). 

Once the material filling the nearby regions start to fall towards the black hole, its angular momentum makes it to move along Keplerian orbits preventing it from reaching the
event horizon. To allow the material to get closer to the black hole, the angular momentum must be redistributed, which requires some release of kinetic energy that can be done
via viscous friction (e.g.  Shakura \& Sunyaev 1973; see also Lynden-Bell \& Pringle 1974) or some kind of wind or collimated outflow/jet (e.g. Bogovalov \& Kelner 2008). In this
way, matter can reach the innermost part close to the horizon. 

In the context of dissipative accretion disks, the magneto rotational instability (MRI; see, e.g., Balbus \& Hawley 1998) is presently the best candidate to yield the conditions
for efficient accretion and subsequent magnetic dominated jet formation and collimation (e.g. Beckwith et al. 2008). The MRI mechanism would play the role of the phenomenological
viscosity introduced by Shakura \& Sunyaev (1973), and would lead to the formation of the ordered magnetic fields in the inner accretion disk regions required for jet launching
(e.g. Kato et al. 2004; Fragile 2008). Concretely, for the case of a rotating black hole, the amplified magnetic field gets strongly bent by the material moving in the ergosphere,
generating Alfven waves that carry angular momentum and energy taken from the rotational energy of the compact object (e.g. Koide et al. 2002; Barkov \& Komissarov 2008), although
differential rotation of the accretion disk alone can also lead to jet production (e.g. Lynden-Bell 1996). 

\subsection{Outflows and particle acceleration}\label{outf}

Supersonic jets are observationally and theoretically linked to the production of non-thermal emission. The kinetic energy associated with the jet matter motion can be converted
into non-thermal particle energy via acceleration through different mechanisms. The most common mechanism in many astrophysical situations is diffusive or Fermi-I shock
acceleration (e.g. Bell 1978a, 1978b; Drury 1983), which would take place in AGNs, GRBs and MQs for instance in the internal shock scenario for the production of high-energy
emission (e.g. Rees 1978; Kobayashi et al. 1997; Kaiser et al. 2000), and also at the shocks produced when jets terminate in the external medium (e.g. Blandford \& Rees 1974;
M\'esz\'aros \& Rees 1997; Heinz \& Sunyaev 2002). Another candidate for accelerating particles in jet regions where the magnetic field is relatively high and turbulent is the
Fermi-II type acceleration mechanism (e.g. Fermi 1949), and under the presence of significant velocity gradients, shear acceleration (e.g. Berezhko \& Krymskii 1981; Rieger \&
Duffy 2004). Fermi-II and shear acceleration may be behind the emission at intermediate scales, in regions where faint and extended radiation is produced. A discussion of the
occurrence of these three acceleration processes in AGNs, GRBs, and MQs can be found in Rieger et al. (2007). 

In general, the acceleration of particles requires two mediums with relative velocity (or one medium with a velocity gradient), and/or some sort of scattering centers (e.g.
magnetic inhomogeneities) moving isotropically in each medium. Particles cross the separation, a shock or velocity gradient region, between the two mediums, as in the case of
Fermi-I and shear acceleration, and/or get scattered in the moving centers, as in the case of Fermi-II acceleration. In these two ways, particles gain energy with each
forth-and-back shock crossing, and with the accumulative effect of many center scatterings. We note that the scattering centers are required also for Fermi-I acceleration since
particles have to cross the discontinuity several times. For this, particles are deflected by these centers and redirected towards this surface after each crossing. Actually, as
discussed in Rieger et al. (2007), the three mechanisms discussed above, Fermi-I, Fermi-II and shear, could compete in the same source.

When very fast outflows are involved, and under very dense radiation fields, the converter mechanism can be efficient for electrons, and protons (e.g. Derishev et al. 2003; Stern
\& Poutanen 2006). In this mechanism, particles suffer interactions with the photons of the intense radiation field and channel most of their energy to a neutral particle, a
photon (for electrons) or a neutron (for protons). These conversions can allow acceleration to become extremely efficient, challenging even the classical electrodynamical limit
(Hillas 1984). Very close to the black-hole horizon, particles could be also accelerated by magnetocentrifugal forces, like in the pulsar magnetosphere (e.g. Neronov \& Aharonian
2007; Rieger \& Aharonian 2008). Finally, magnetic reconnection (e.g. Romanova \& Lovelace 1992; Zenitani et al. 2001) could also accelerate particles in some specific situations,
namely in the regions where the magnetic field is expected to be dynamically dominant, e.g. in the 
base of the jet. In this process, field lines of opposite polarity that get
very close suddenly connect to each other, heating the plasma and accelerating particles.

\subsection{Types of sources}

The main source types that harbor black holes and produce non-thermal emission are: AGNs, which are galaxies harboring an actively accreting supermassive ($\sim
10^6-10^{10}$~M$_{\odot}$) black hole and can produce non-thermal emitting relativistic outflows (e.g. Begelman et al. 1984; Rees 1984; Osterbrock 1993); GRBs (e.g. M\'esz\'aros
2006), which would result from the collapse of a very massive star or from the coalescence of two neutron stars or a neutron star and a black hole (e.g. Eichler et al. 1989;
Woosley 1993; Paczynski 1998), yielding long and short GRBs (Kouveliotou et al. 1993); and MQs, which are X-ray binaries with relativistic jets (e.g. Mirabel \& Rodr\'iguez 1999;
Rib\'o 2005; Fender 2006). 

In all these three kinds of object, AGNs, GRBs, and MQs, the site of non-thermal emission is generally the jet or the jet termination regions, although there are some cases in
which the radiation could also be generated outside the jet, like in the accretion disk corona (e.g. Lynden-Bell 1969; Pineault 1982; Torricelli-Ciamponi \& Pietrini 2005), or the
jet surroundings, like the stellar wind in high-mass MQs (e.g. Bosch-Ramon et al. 2008). The dominant non-thermal processes are\footnote{Note the reader that the overview of
high-energy processes presented here is not exhaustive.} synchrotron and inverse Compton (IC) emission for leptons (see, e.g., Blumenthal \& Gould 1970), and proton-proton ($pp$)
collisions (e.g. Kelner, Aharonian, \& Bugayov 2006), photo-meson production (e.g. Kelner \& Aharonian 2008), and photodisintegration (e.g. Anchordoqui et al. 2007) for hadrons.

Synchrotron radiation is produced by electrons that move in a magnetized medium and suffer Lorentz forces spiraling around the magnetic field lines. IC photons are
produced when ambient photons are scattered by electrons getting a large increase of their energy and momentum. $pp$ interactions, and photo-meson production resulting from
proton-photon collisions, produce photons mainly via the production of neutral and charged pions that decay to gamma-rays, neutrinos, and charged muons/electron-positron pairs,
which also emit photons. Finally, the disintegration of heavy nuclei by collisions with photons also lead to gamma-ray production. Further discussion of these processes in the
context of MQs can be found in Bosch-Ramon \& Khangulyan (2008). It is worth remarking that for environments with large matter densities, suitable for efficient gamma-ray
production from $pp$ interactions, relativistic Bremsstrahlung produced by electrons moving in the electric field of ions (e.g. Blumenthal \& Gould 1970) could also be an
efficient process, although generally it will be overcome by synchrotron and IC scattering.

As explained in Sect.~\ref{outf}, the jet base, the intermediate scale region, and the jet termination point, are the main regions where radiation is produced (basically along the
whole jet), or expected to be produced when no direct spatial information is at hand, in sources harboring black holes and producing collimated outflows. The different processes
that lead to non-thermal emission generate photons in wide bands of the electromagnetic spectrum. In the case of synchrotron, the emission spans from radio to X-rays and sometimes
soft gamma-rays, depending on the electron energy and the magnetic field. In the case of IC, scattered photons can generally go from X-rays up to very high energies, although
lower and higher energies are also possible. In hadronic processes, photons are produced in the gamma-ray range, and neutrinos and secondary pairs with similar energies to those
of gamma-rays can also be generated. These secondary leptons will also radiate synchrotron and IC in wide energy ranges. For works regarding models of emission for different
regions, mechanisms and spatial scales, in AGNs, GRBs and MQs, we refer to the literature (a non exhaustive list of works and reviews: Jones et al. 1974; Ghisellini et al. 1985;
Dermer et al. 1992; Sikora et al. 1994; Atoyan \& Aharonian 1999; Aharonian 2002; Aharonian et al. 2006; Levinson 2006; M\'eszaros 2006; B\"ottcher 2007; Romero 2008b; Paredes
2008b; Bosch-Ramon \& Khangulyan 2008; and references therein). Finally, the relativistic emitting particles themselves may escape the source and eventually reach the Earth, being
detected as CRs. MQs in the low energy part but hardly at dominant levels (e.g. Heinz \& Sunyaev 2002), and GRBs and AGNs at ultra high energies (e.g. Hillas 1984), are all
expected contributors to the CR spectrum that reaches the
Greisen-Zatsepin-Kuzmin (GZK) cut-off.

It is still worthy to note that in the innermost regions of jets, and in the surroundings of jets with nearby powerful photon sources, the absorption of gamma-rays by
photon-photon interactions (e.g. Gould \& Schr\'eder 1967) will lead to the creation of secondary pairs that may, either via synchrotron or IC interactions, release again the
absorbed energy. In case the magnetic field energy density were much smaller than the photon field energy density, efficient photon-photon cascading could take place (e.g.
Akharonian \& Vardanian 1985; Bednarek 1997; Coppi \& Aharonian 1997; Orellana et al. 2007; Khangulyan et al. 2008, etc.).

\section{Discussion}

\subsection{Differences and similarities between source types}

The different objects harboring a black hole and producing non-thermal radiation share the same basic elements: the black hole itself; an accretion disk; the presence of a jet or
outflow; particle acceleration; and the candidates to dominant radiative processes discussed in previous section. Nevertheless, despite being so similar, there are still important
differences between the black-hole powered sources beyond those obvious, i.e. the black-hole mass (and thereby size) and spin\footnote{The charge is generally not considered in
astrophysical black holes due to the short time in which they discharge by opposite charge accretion.}, the distance to us, and the accretion rate. There are as well the
environmental factors. For instance, the presence of a external source of photons, its location with respect to the jet, its luminosity and the energy of the photons, all can
determine the spectrum at high and very high energies via IC, photo-meson production, photo-disintegration, or photon-photon absorption. The external medium density and velocity
can be also very important regarding radiative processes like gamma-ray production from $pp$ interactions, as well as for the jet propagation and stability. For instance, in the
case of AGNs, different powers and medium densities give rise to different jet termination structures (e.g. Fanaroff \& Riley 1974; for MQ jet-medium interactions, see e.g. Bordas
et al. 2008 and references therein). Also, GRBs can show very different phenomena during their afterglow phase, when the jet terminates, depending on the characteristics of the
environment, since it may be determined by a massive star wind for a collapsar long GRB, or could be a much less dense medium in the case of coalescence of two compact objects
(Chevalier \& Li 1999). In the case of MQs with a massive stellar companion, which fills the jet surroundings with powerful winds, hydrodynamical and non-thermal phenomena can
take place given the presence of a magnetized and moving plasma colliding with the jet (e.g. Perucho \& Bosch-Ramon 2008; Bosch-Ramon et al. 2008; 
Araudo, Bosch-Ramon \& Romero 2009).

\subsection{Prospects}

Our present knowledge on the processes taking place in the black-hole vicinity, as well as in the outflows generated from there, is limited to the sensitivity, angular and energy
resolution of the instruments working in the whole spectral range, mainly in radio, X-rays, and gamma-rays. Nevertheless, the prospects of the new instrumentation in all these
ranges, with significant improvements in sensitivity and energy and angular resolution, are for the near future very important observational discoveries and thereby advances in
our theoretical knowledge. For instance, VLBI interferometric techniques can unveil the closest regions to the black-hole horizon; polarization studies can say much about the role
of magnetic fields in the jet formation, collimation and evolution; radio and X-ray imaging can answer fundamental questions on the jet structure at different scales; and
gamma-ray observations can unveil the physical processes taking place in plasmas of very extreme conditions, like in the the jet base, in relativistic shell collisions,
or in interactions with the environment.

\agradecimientos 
The author thanks Gustavo E. Romero for useful comments on the manuscript. 
The author wants also to thank  the organizers of the meeting for giving him the opportunity to give this lecture 
in the 51st meeting of the Argentinian Astronomical Association. 
V.B-R. gratefully acknowledges support
from the Alexander von Humboldt Foundation. V.B-R. acknowledges support by DGI of MEC under grant AYA2007-68034-C03-01, as
well as partial support by the European Regional Development Fund (ERDF/FEDER).

\begin{referencias}

\vskip 1cm
\reference Abraham, J., Abreu, P., Aglietta, M. 2007, Science, 318, 938
\reference Aharonian, F.~A. 2002, MNRAS, 332, 215 
\reference Aharonian, F.~A., Anchordoqui, L.~A., Khangulyan, D. \&, Montaruli, T. 2006, J. Phys.Conf.Ser., 39, 408 
\reference Akharonian, F. A., Vardanian, V. V., 1985, Ap\&SS, 115, 31 
\reference Anchordoqui, L.~A., Beacom, J.~F., Goldberg, H., Palomares-Ruiz, S., \& Weiler, T.~J. 2007, Phys. Rev. Lett., 98, 1101
\reference Araudo, A.~T. Bosch-Ramon, V. \& Romero, G.~E. 2009, A\&A, submitted	
\reference Atoyan, A.~M. \& Aharonian, F.~A. 1999, MNRAS, 302, 253
\reference Beckwith, K., Hawley, J.~F., Krolik, J.~H. 2008, ApJ, 678, 1180 
\reference Balbus, S.~A. \& Hawley, J.~F. 1998, Rev. Mod. Phys., 70, 1
\reference Barkov, M.~V. \& Komissarov, S.~S. 2008, MNRAS, 385, 28
\reference Bednarek, W. 1997, A\&A, 322, 523 
\reference Begelman, M.~C., Blandford, R.~D., \& Rees, M.~J. 1984, Rev. Mod. Phys., 56, 255	
\reference Bell, A. R. 1978, MNRAS, 182, 147 
\reference Bell, A. R. 1978, MNRAS, 182, 443 
\reference Berezhko, E.~G. \& Krymskii, G.~F. 1981, Sov. Astron. Lett., 7, 352
\reference Blandford, R.~D. 1976, MNRAS, 176, 465
\reference Blandford, R.~D. \& Rees, M.~J. 1974, MNRAS, 169, 395 
\reference Blandford, R.~D. \& Znajek, R.~L. 1977, MNRAS, 179, 433 
\reference Blandford, R.~D. \& Payne, D. G. 1982, MNRAS, 199, 883 
\reference Blumenthal, G.~R. \& Gould, R.~J. 1970, Rev. Mod. Phys., 42, 237
\reference Bogovalov, S. \& Kelner, S. 2008, MNRAS, submitted [astro-ph/0809.0429]
\reference B\"ottcher, M. 2007, Ap\&SS, 309, 95
\reference Bolton, C.~T. 1972, Nature, 235, 271
\reference Bordas, P., Bosch-Ramon, V., Paredes, J.~M., \& Perucho, M. 2008, contribution for VII Microquasar Workshop, Foca 2008
\reference Bosch-Ramon, V. \& Khangulyan, D. 2008, Int. J. Mod. Phys.~D, in press [astro-ph/0805.4123]
\reference Bosch-Ramon, V., Khangulyan, D., \& Aharonian, F.~A. 2008, A\&A, 482, 397 
\reference Burbidge, G.~R. 1956, ApJ, 124, 416
\reference Casares J. 2001, LNP, 563, 277 	
\reference Chevalier, R.~A. \& Li, Z. 1999, ApJ, 520, 29
\reference Coppi, P.~S. \& Aharonian, F.~A. 1997, ApJ, 487, 9
\reference Derishev, E.~V., Aharonian, F.~A., Kocharovsky, V.~V., \& Kocharovsky, Vl.~V. 2003, Phys. Rev. D, 68, 3003
\reference Dermer, C. D., Schlickeiser, R., \& Mastichiadis, A. 1992, A\&A, 256, 27
\reference Drury, L.~O., 1983, Rep. Prog. Phys., 46, 973 
\reference Eichler, D., Livio, M., Piran, T., \& Schramm, D.~N. 1989, Nature, 340, 126
\reference Einstein, A. 1915, Preussische Akademie der Wissenschaften, p.844
\reference Fanaroff, B.~L. \& Riley, J.~M. 1974, MNRAS, 167, 31 
\reference Fender, R. 2006, in: Compact stellar X-ray sources. Ed. Lewin, W. \& van der Klis, M.. Cambridge Astrophysics Series, 39, 381
\reference Fermi, E. 1949, Phys. Rev., 75, 1169
\reference Fragile, P.~C. 2008, Proceedings of Science, invited review for VII Microquasar Workshop, Foca 2008 [astro-ph/0810.0526]	
\reference Ghisellini, G., Maraschi, L., \& Treves, A. 1985, A\&A, 146, 204
\reference Gould, R.~J. \& Schr\'eder, G.~P. 1967, Phys. Rev, 155, 1404
\reference Heinz, S. \& Sunyaev, R. 2002, A\&A, 390, 751
\reference Hillas, A.~M. 1984, ARA\&A, 22, 425 
\reference Kaiser, C.~R., Sunyaev, R., \& Spruit, H.~C. 2000, A\&A, 356, 975 
\reference Kato, Y., Mineshige, S., \& Shibata, K. 2004, ApJ, 605, 307 
\reference Kelner, S.~R., Aharonian, F.~A., \& Bugayov, V.~V. 2006, Phys. Rev. D, 74, 4018
\reference Kelner, S.~R. \& Aharonian, F.~A. 2008, Phys. Rev. D, 78, 4013
\reference Kerr, R.~P. 1963, Phys. Rev. Lett., 11, 237
\reference Khangulyan, D., Aharonian, F., \& Bosch-Ramon, V. 2008, MNRAS, 383, 467 
\reference Kobayashi, S., Piran, T., \& Sari, R. 1997, ApJ, 490, 92 
\reference Koide, S., Shibata, K., Kudoh, T., Meier, D.~L. 2002, Science, 295, 1688
\reference Kouveliotou, C., Meegan, C.~A., \& Fishman, G.~J. 1993, ApJ, 413, 101
\reference Levinson, A. 2006, Int. J. Mod. Phys.~A, 21, 30
\reference Lovelace, R.~V.~E. 1976, Nature, 262, 649
\reference Lynden-Bell, D. 1969, Nature, 223, 690
\reference Lynden-Bell, D. 1996, MNRAS, 279, 389 
\reference Lynden-Bell, D. \& Pringle, J.~E. 1974, MNRAS, 168, 603 
\reference M\'esz\'aros, 2006, Rep. Prog. Phys., 69, 2259
\reference M\'esz\'aros, P. \& Rees, M.~J. 1997, ApJ, 476, 232 
\reference Mirabel, I.~F. \& Rodr\'iguez, L.~F. 1999, ARA\&A, 37, 409 		
\reference Neronov, A. \& Aharonian, F.~A. 2007,  ApJ, 671, 85
\reference Newmann, E.~T., Couch, R., Chinnapared, K., et al. 1965, J. Math. Phys. 6, 918
\reference Nordstr\"om, G., 1918, Verhandl. Koninkl. Ned. Akad. Wetenschap., Afdel. Natuurk., Amsterdam, 26, 1201 
\reference Orellana, M., Bordas, P., Bosch-Ramon, V., Romero, G.~E.,\&  Paredes, J.~M. 2007, A\&A, 476, 9 
\reference Osterbrock, D.~E. 1993, ApJ, 404, 551 	
\reference Paczynski, B. 1998, ApJ, 494, 45
\reference Paredes, J.~M. 2008a, Lecture Notes from the First La Plata International School on Astronomy and Geophysics
\reference Paredes, J.~M. 2008b, Int. Jour. Mod. Phys. D, 17, 1849 
\reference Penrose, R. \& Floyd, G.~R. 1971, Nature, 229, 177
\reference Perucho, M. \& Bosch-Ramon, V. 2008, A\&A, 482, 917 
\reference Pineault, S. 1982, A\&A, 109, 294 
\reference Punsly, B. \& Coroniti, F.~V. 1990, ApJ, 350, 518
\reference Rees, M.~J. 1978, MNRAS, 184, 61 
\reference Rees, M.~J. 1984, ARA\&A, 22, 471 
\reference Reissner, H., 1916, Annalen der Physik, 50, 106
\reference Rib\'o, M. 2005, ASPC, 340, 269
\reference Rieger, F.~M. \& Duffy, P. 2004, ApJ, 617, 155
\reference Rieger, F. \& Aharonian, F.~A. 2008, A\&A, 479, 5 
\reference Rieger, F.~M., Bosch-Ramon, V., \& Duffy, P. 2007, Ap\&SS, 309, 119
\reference Romanova, M.~M., Lovelace, R.~V.~E. 1992, A\&A, 262, 26
\reference Romero, G.~E. 2008a, Lecture Notes from the First La Plata International School on Astronomy and Geophysics [astro-ph/0805.2082]
\reference Romero, G.~E. 2008b, Proceedings of Science, contribution for VII Microquasar Workshop, Foca 2008 [astro-ph/0810.0202]
\reference Romero, G.~E., Combi, J.~A., Perez Bergliaffa, S.~E., \& Anchordoqui, L.~A. 1996, Astropart. Phys., 5, 279
\reference Salpeter, E.~E. 1964, ApJ, 140, 796 
\reference Sch\"odel, R., Ott, T., Genzel, R., et al. 2002, Nature, 419, 694	
\reference Schwarzschild, K., 1916a, Sitzungsberichte der Königlich Preussischen Akademie der Wissenschaften 1, 189
\reference Schwarzschild, K., 1916b, Sitzungsberichte der Königlich Preussischen Akademie der Wissenschaften 1, 424
\reference Shakura, N.~I. \& Syunyaev, R.~A. 1973, A\&A, 24, 337
\reference Sikora, M., Begelman, M.~C., \& Rees, M.~J. 1994, ApJ, 421, 153
\reference Stern, B.~E. \& Poutanen, J. 2006, MNRAS, 372, 1217
\reference Jones, T.~W., O'dell, S.~L., \& Stein, W.~A. 1974, 
\reference Torricelli-Ciamponi, G., Pietrini, P., \& Orr, A. 2005, A\&A, 438, 55 
\reference Webster, B.~L. \& Murdin, P. 1972, Nature, 235, 37
\reference Woosley, S.~E. 1993, ApJ, 405, 273	
\reference Zenitani, S. \& Hoshino, M. 2001, ApJ, 562, 63

\end{referencias}

\end{document}